\documentclass[final,1p,times]{elsarticle} 
\usepackage{graphicx} 
\usepackage{amssymb} 
\usepackage{amsthm} 
\usepackage{lineno} 

\def\sqrts{$\sqrt{s}$ }
\def\pt{$p_T$ }
\def\pT{$p_T$}
\def\mT{$m_T$}
\def\mt{$m_T$ }
\def\ee{$e^+e^-$ }

\def\gev2{GeV/$c^2$} 

\def\mev2{MeV/$c^2$}

\def\AuAu{Au+Au }
\def\CuCu{Cu+Cu }
\def\pp{p+p }
\def\dAu{d+Au }
\def\teff{$T_{eff}$ }

\journal{Nuclear Physics A} 
\begin{document} 

\begin{frontmatter} 


\title{Dileptons And Photons At RHIC Energies }

\author{Axel Drees}

\address{Stony Brook University, Stony Brook, NY 11777, USA}

\begin{abstract} 
Comparing dilepton data from \AuAu to \pp collisions at RHIC energies 
PHENIX revealed two striking features: (i) a large excess of the 
dilepton yield at low mass and low \pt in central \AuAu and (ii) a 
significant enhancement of direct real photon radiation. Both features
can be interpreted as thermal radiation from the hot quark matter produced
at RHIC. Unlike at lower beam energies, at RHIC the contribution from 
the hot hadronic phase alone seems insufficient to account for the data. 
This suggests large a contribution from a partonic phase. Both experimental 
and theoretical progress are required for more detailed conclusions.
  
\end{abstract} 

\end{frontmatter} 



\section{Introduction}\label{sec:intro}

Electromagnetic radiation is regarded as an ideal tool to probe 
strongly interacting matter \cite{shuryak79}. Experiments at SPS
produced a wealth of data and inspired much theoretical work that  
advanced our understanding of meson properties in matter and the relation 
to chiral symmetry restoration \cite{spsreview,theoryreview}. 
The PHENIX experiment continues the experimental investigation 
at RHIC in a new energy domain. In this review I
report on the recent PHENIX results and discuss the comparison to model 
calculations. 

\begin{figure}[h!]
  \vspace{-0.2cm}
  \begin{minipage}{1.\linewidth}
  \begin{center}
      \includegraphics[width=0.49\textwidth]{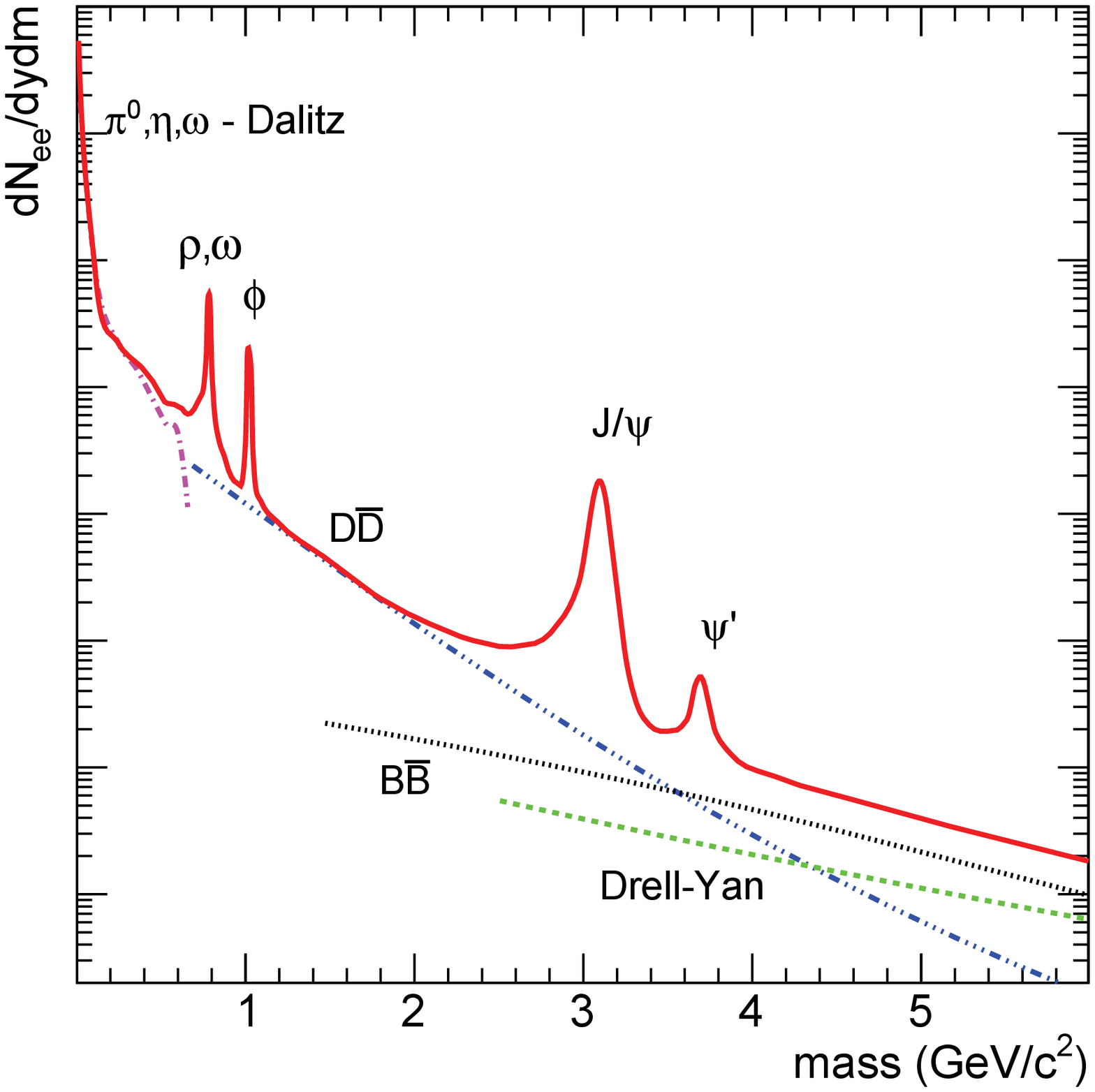}
      \includegraphics[width=0.45\textwidth]{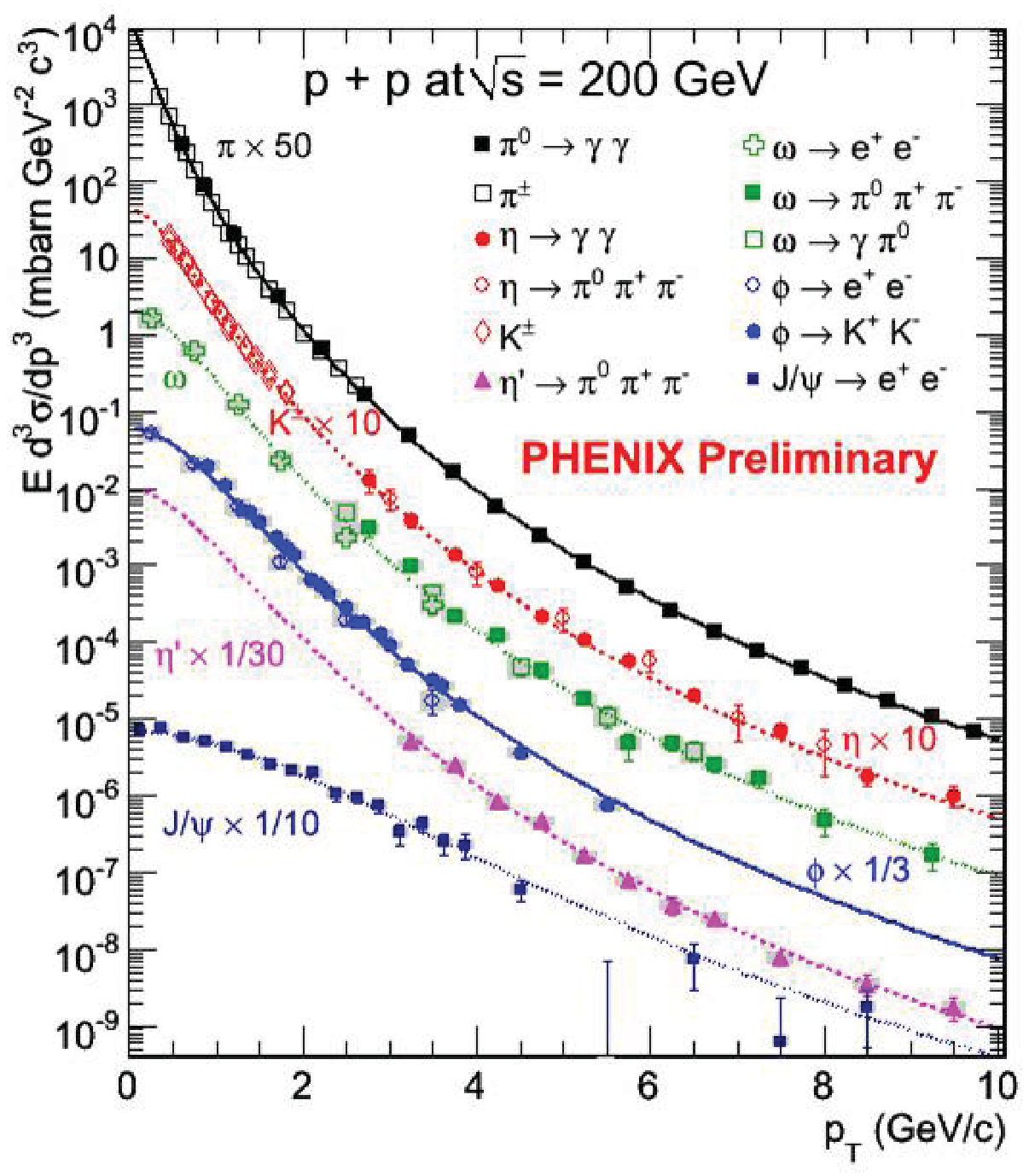}
  \end{center}
  \end{minipage}

  \vspace{-0.7cm}
  \begin{flushleft}
  \begin{minipage}{.48\linewidth}
      \caption{\label{fig:dilepton-cartoon} Schematic view of the 
        \ee pair mass distribution from p+p collisions 
        at 200 GeV.  }
  \end{minipage}
  \end{flushleft}
  \vspace{-1.65cm} 
  \begin{flushright}
  \begin{minipage}{.48\linewidth}
      \caption{\label{fig:pp-mtscaling} Meson production cross-sections 
        from 200 GeV 
        \pp collisions measured by PHENIX. The data for pseudo-scaler
        and vector mesons are compared to the distributions derived by scaling 
        the shape of the distributions by transverse mass and fitting the 
        overall normalization to data \cite{ppg085}.}  
  \end{minipage}
  \end{flushright}
\vspace{-.2cm}
\end{figure} 

Figure~\ref{fig:dilepton-cartoon} shows a schematic \ee pair mass 
distribution as expected from \pp collisions at RHIC energies. The \ee pairs 
are either from pseudo-scaler or vector-meson decays, typically after 
the collision on the time scale of the strong interaction, or from 
hard-scattering processes like open heavy-flavor production, 
Drell-Yan pair production or quark-gluon ($qg$) Compton scattering, 
all occurring early in the collision. In addition to these sources
quark matter produced in heavy ion collisions may emit 
thermal radiation, leave imprints from medium modifications 
of the meson properties, and modify the contribution from 
hard-scattering processes.  

Isolating the medium effects from the expected sources is the experimental 
challenge and this challenge is two fold. First one needs to determine and 
subtract the pair background, which in central \AuAu collisions is more 
than a factor of hundred larger than the signal. This subtraction is very 
involved and dominates the systematic uncertainties. After many years of 
analysis effort this subtraction is well under control 
(see \cite{ppg075,ppg088} for details). The second challenge is 
to accurately determine the expected
sources shown schematically in Fig.~\ref{fig:dilepton-cartoon}. PHENIX was able
to determine these expected sources from data on 
meson production 
cross sections (see Fig.~\ref{fig:pp-mtscaling} 
and \cite{ppg085}) and on heavy-flavor production \cite{ppg065,ppg066}, 
measured in the same experiment.

\section{Reference data from \pp collisions}\label{sec:pp}

\begin{figure}[h!]
  \vspace{-0.5cm}
  \begin{minipage}{1.\linewidth}
    \begin{center}
      \includegraphics[width=0.7\textwidth]{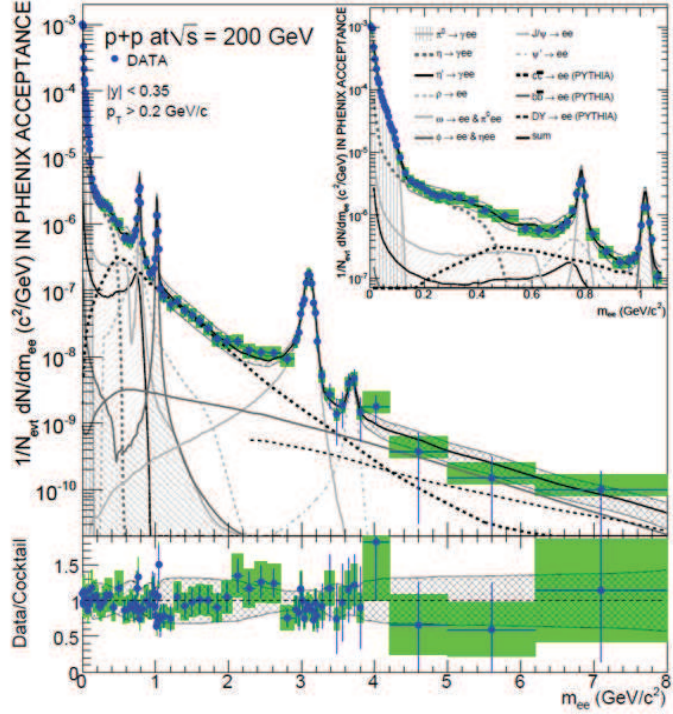}

    \end{center}
  \end{minipage}

  \vspace{-0.3cm}
    \begin{minipage}{\linewidth}
      \caption{\label{fig:pp-mass} Electron-positron pair data from
        200 GeV \pp collisions observed by the PHENIX experiment 
        \cite{ppg085}. The
        solid line depicts the contribution from hadron decays estimated
        on the basis of the most precise available data on neutral meson
        production and charm/bottom production. Data are corrected 
        for efficiency and require both electron and positron in the detector 
        acceptance.}
    \end{minipage}
  \vspace{-0.3cm}
\end{figure}

Data from \pp collisions are used to test the analysis method and serves as 
precision reference for data from heavy ion collisions. The data are 
published in \cite{ppg085}. They are compared to the expected 
yield in Fig.~\ref{fig:pp-mass}. It is evident from the ratio 
data-to-expected given in the lower panel that the data are 
in excellent agreement with the expectation within the systematic 
uncertainties of about 30\%. 

The data are very precise and can be used to independently measure 
some of the expected contributions. This was done for the vector 
mesons $\omega, \phi$ \cite{dahms} and $J/\Psi$ \cite{ppg069} 
as well as for heavy-flavor production \cite{ppg085} and 
$qg$ Compton scattering. The later contribution can be isolated in the 
kinematic region where the mass is smaller than the transverse momentum 
\cite{ppg086,Akiba}. The ratio of direct-to-inclusive virtual photons can 
then be extrapolated to mass equal zero and compared to perturbative 
QCD calculations \cite{vogelsang}. As shown in Fig.~\ref{fig:pp-rgamma} 
the agreement is very good.     

\section{Data from \AuAu collisions}\label{sec:AuAu}

\begin{figure}[h!]

  \vspace{-.1cm}
  \begin{minipage}{1.0\linewidth}
    \begin{center}
      \includegraphics[width=0.39\textwidth]{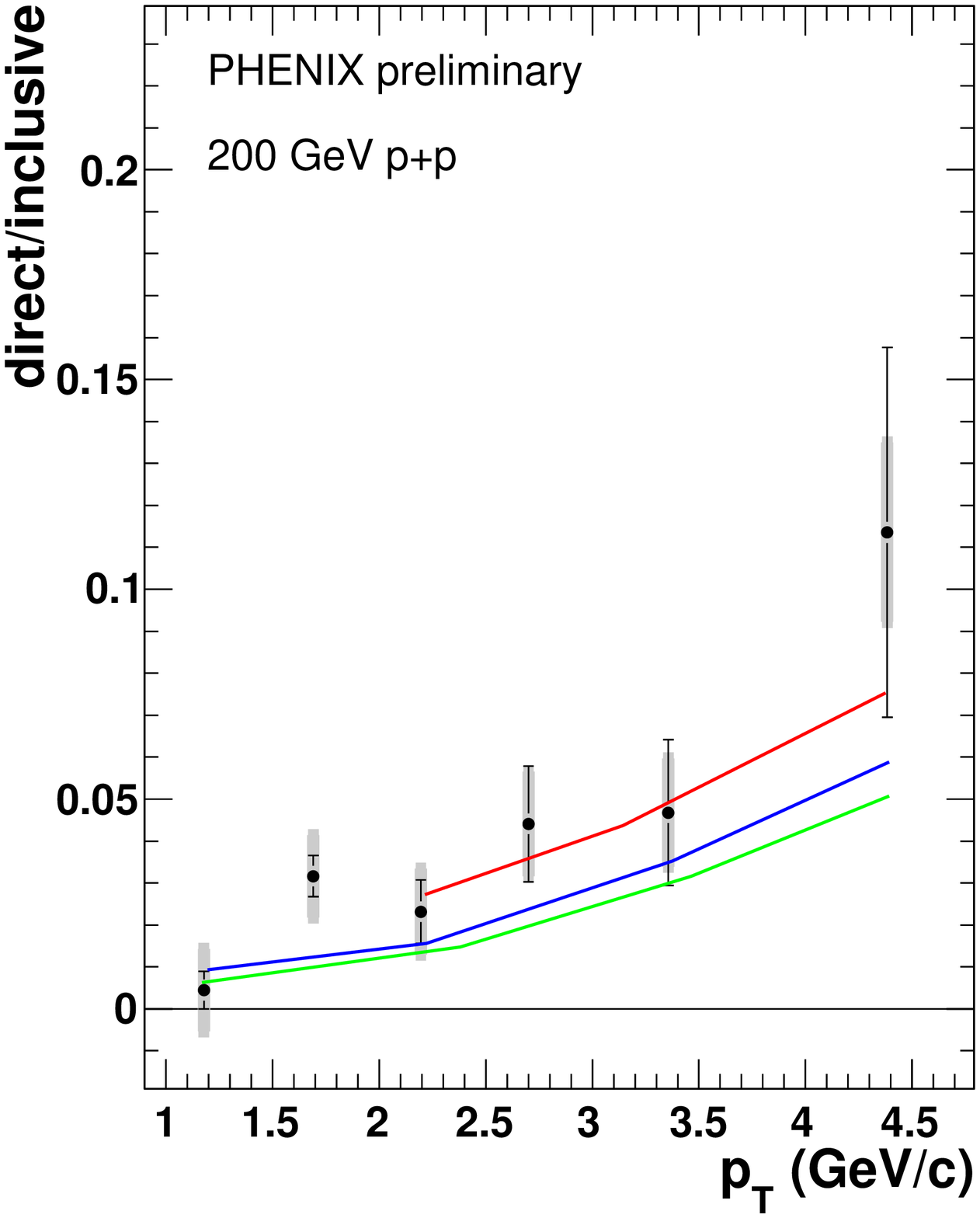}
      \includegraphics[width=0.59\textwidth]{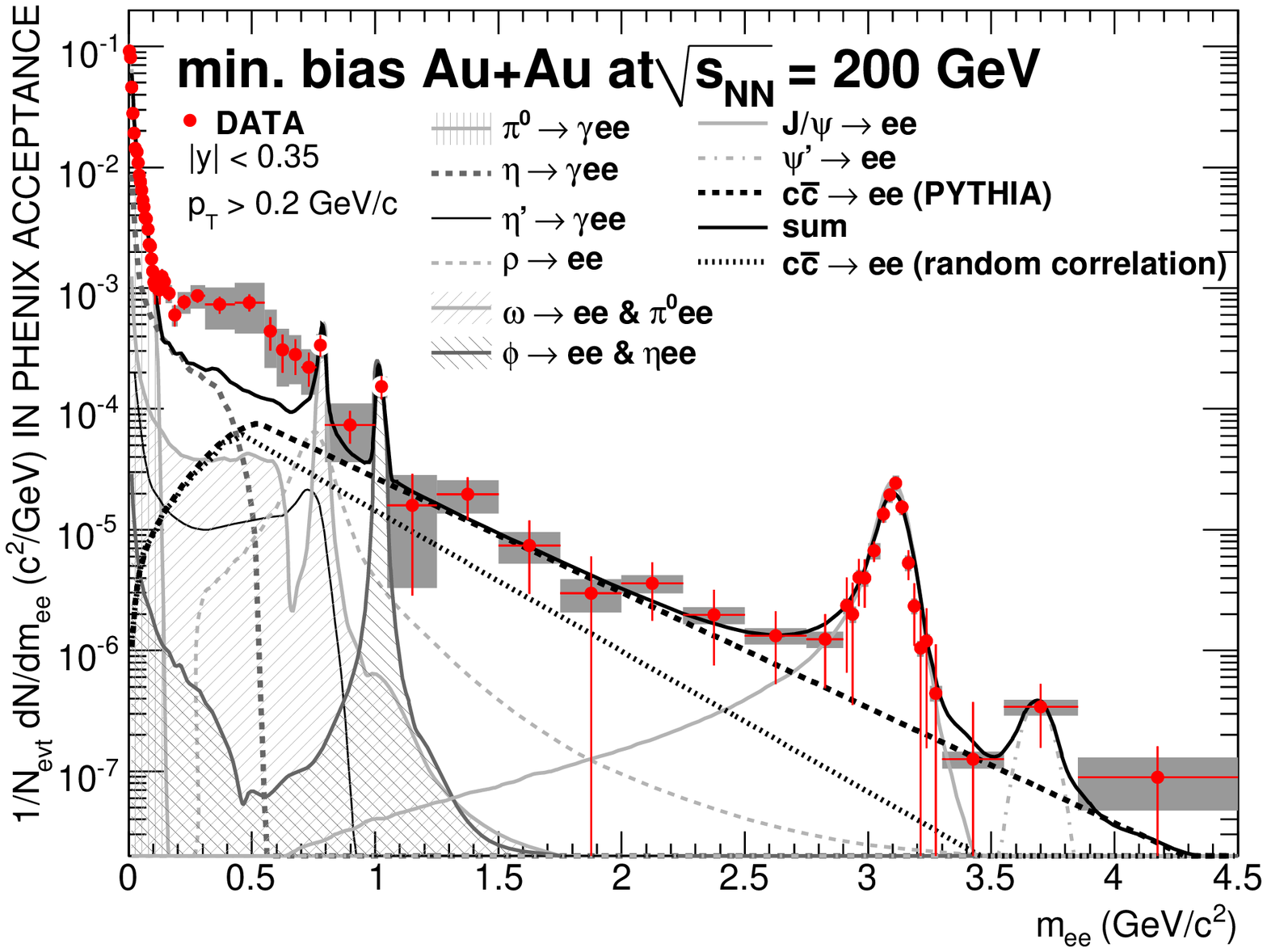}
    \end{center}
  \end{minipage}

  \vspace{-.3cm} 
  \begin{flushleft}
    \begin{minipage}{.39\linewidth}
      \caption{\label{fig:pp-rgamma} Ratio of direct-to-inclusive dilepton 
        pairs in the kinematic range $m << p_T$. The data are compared to 
        perturbative QCD calculations for direct photon production 
        \cite{vogelsang}.}
    \end{minipage}
  \end{flushleft}
  \vspace{-2.37cm} 
  \begin{flushright}
    \begin{minipage}{.58\linewidth}
      \caption{\label{fig:AuAu-mass} Inclusive \ee  mass spectrum
        from \AuAu collisions at \sqrts = 200 GeV measured by PHENIX. 
        The presentation 
        is identical to that in Fig.~\ref{fig:pp-mass}. The
        statistical errors are shown as bars and the systematic errors 
        are marked independently by a band around each data point. }
    \end{minipage}
  \end{flushright}
  \vspace{-.2cm}
\end{figure}

With our understanding of the expected sources we have a solid base 
line to search for effects off the medium. Already from a first 
glance at the mass distribution from \AuAu collisions 
(Fig.~\ref{fig:AuAu-mass}) one finds a striking enhancement of the 
data above the expected yield that was established in \pp collisions. 
In the mass range from 150 to 750 \mev2 the yield is a 
factor $3.4\pm 0.2(stat.)\pm 1.3(sys.)\pm 0.7(model)$ above the expectation. 
     
\begin{figure}
  \vspace{0.1cm}
  \begin{minipage}{1.\linewidth}
  \begin{center}
      \includegraphics[width=0.49\textwidth]{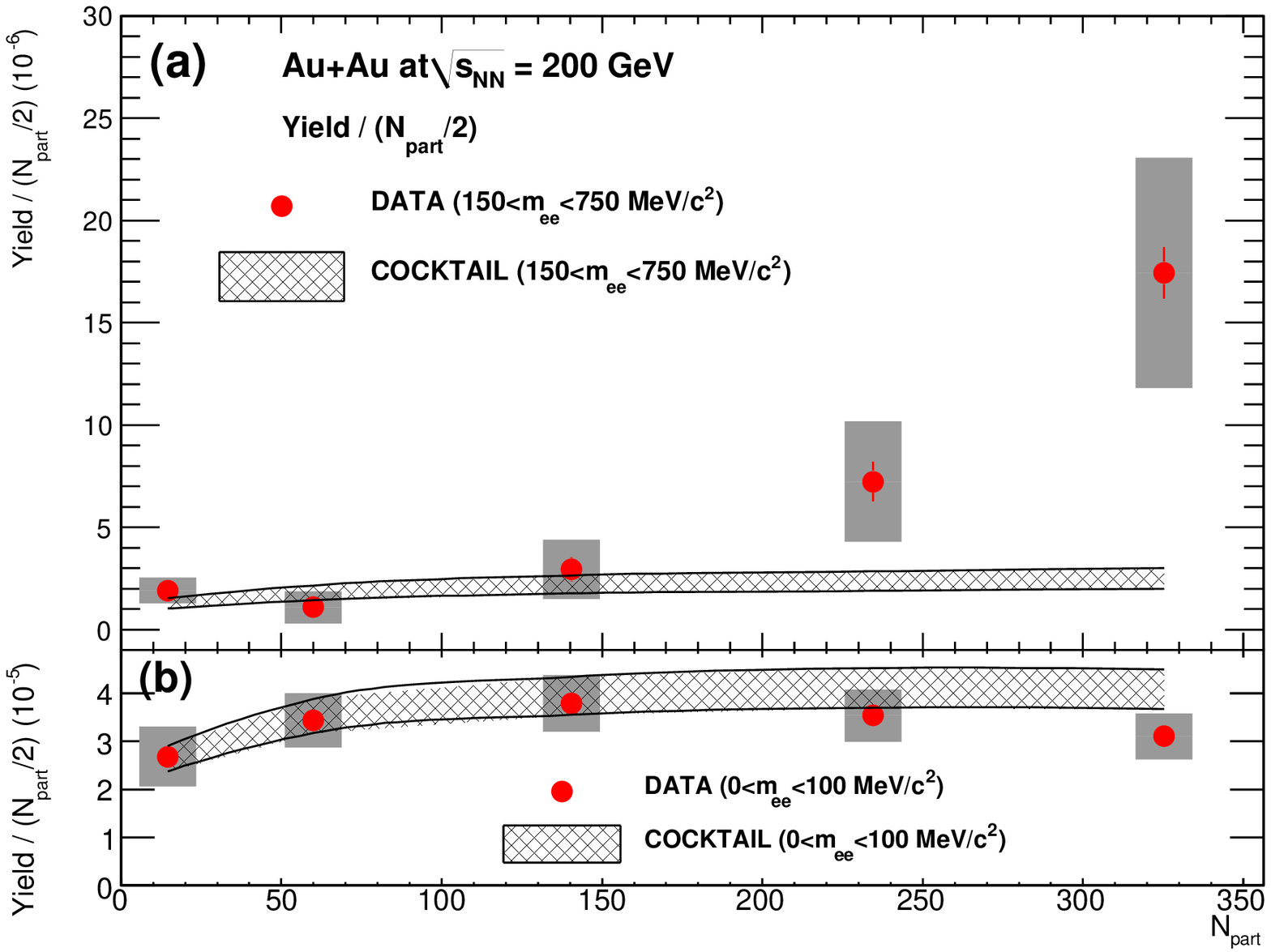}
      \includegraphics[width=0.5\textwidth]{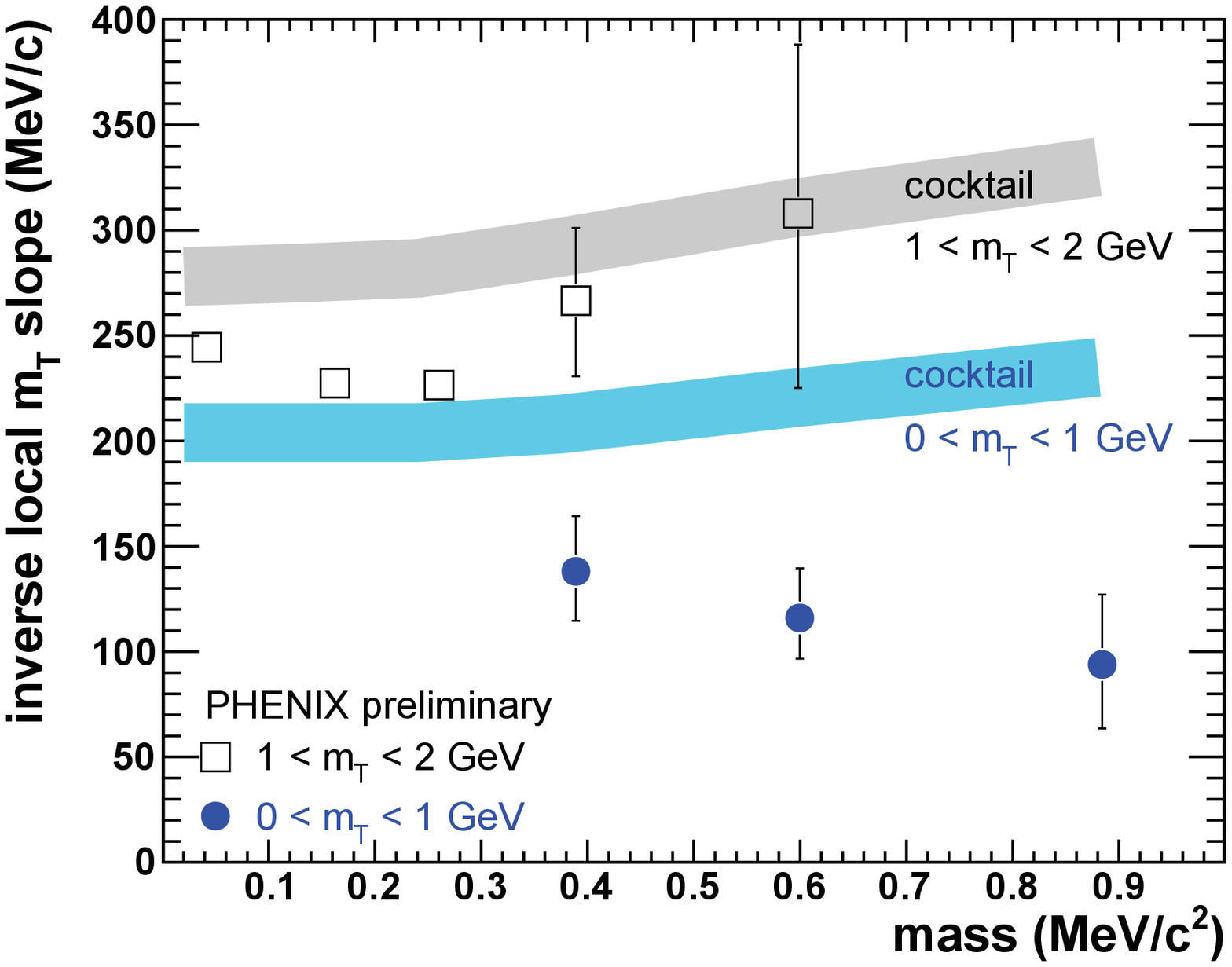}
  \end{center}
  \end{minipage}

  \vspace{-0.2cm}
  \begin{flushleft}
  \begin{minipage}{.48\linewidth}
      \caption{\label{fig:AuAu-centrality} 
        Electron-positron pair yield per number 
        of participants for two mass ranges shown as function of collision 
        centrality. The data are compared to the expectation from known 
        sources. } 
  \end{minipage}
  \end{flushleft}
  \vspace{-2.3cm} 
  \begin{flushright}
  \begin{minipage}{.48\linewidth}
      \caption{\label{fig:AuAu-mtslopes} 
        Inverse slope of the \mt  
        distribution for minimum bias Au+Au collisions as a function of mass
        for two different regions for \mT. Different from previous figures 
        the data were corrected for pair acceptance. Again the data are compared 
        to the expectation from known sources.}
  \end{minipage}
  \end{flushright}
  \end{figure}

In order to investigate the enhancement in more detail one can study its 
centrality and momentum dependence. The centrality dependence of the 
integrated yield is shown in Fig.~\ref{fig:AuAu-centrality} for two mass 
ranges $m<100$ \mev2 and 150 to 750 \mev2, respectively. The yield in the
lower mass bin is dominated by pairs from $\pi^0$ Dalitz-decays and 
consequently follows the centrality dependence of the pion yield. However,
the yield in the higher mass range, which includes the enhancement, 
deviates from the expected dependence and indicates that central collisions 
contribute most of the observed enhancement. The yield from 
$\pi\pi$ or $q\bar{q}$ annihilation in the medium should also increase faster 
than the number of participants, but it is not clear if 
the data can be explained quantitatively.      

Splitting the data in bins of pair \pt reveals that in the mass range 
from 150 to 750 \mev2 the yield is enhanced at all \pT, but not uniformly. The
enhancement is largest at the lowest \pT. To quantify this more
PHENIX has measured the transverse mass (\mT) distributions, corrected for 
pair acceptance, and determined local slopes as function of pair mass. 
The inverse local slope or \teff for the two regions $0<m_T<1$ GeV and 
$1<m_T<2$ GeV is plotted in Fig.~\ref{fig:AuAu-mtslopes} as a function of mass. 
Also shown is \teff expected from hadron decays alone. In general neither 
expectation from hadron decays nor the data show a strong variation of the 
effective temperature over the observed mass range. In the higher 
\mt range the effective  temperatures are similar to the expectation, 280 MeV 
compared to 220 MeV in the data. In the lower 
\mt range the data have a significantly smaller slope of 120 MeV than the
expected 200 MeV. Unfortunately, limited statistics prohibits a more 
detailed analysis. More precise data are needed to subtract the 
hadron decay contribution and isolate the enhancement. However, the effective
temperature of the low \mt component is a rather robust measurement and the 
result will not change much once the excess can be isolated.

  \begin{figure}[h!]
  \vspace{-0.1cm}
  \begin{minipage}{1.\linewidth}
  \begin{center}
      \includegraphics[width=0.45\textwidth]{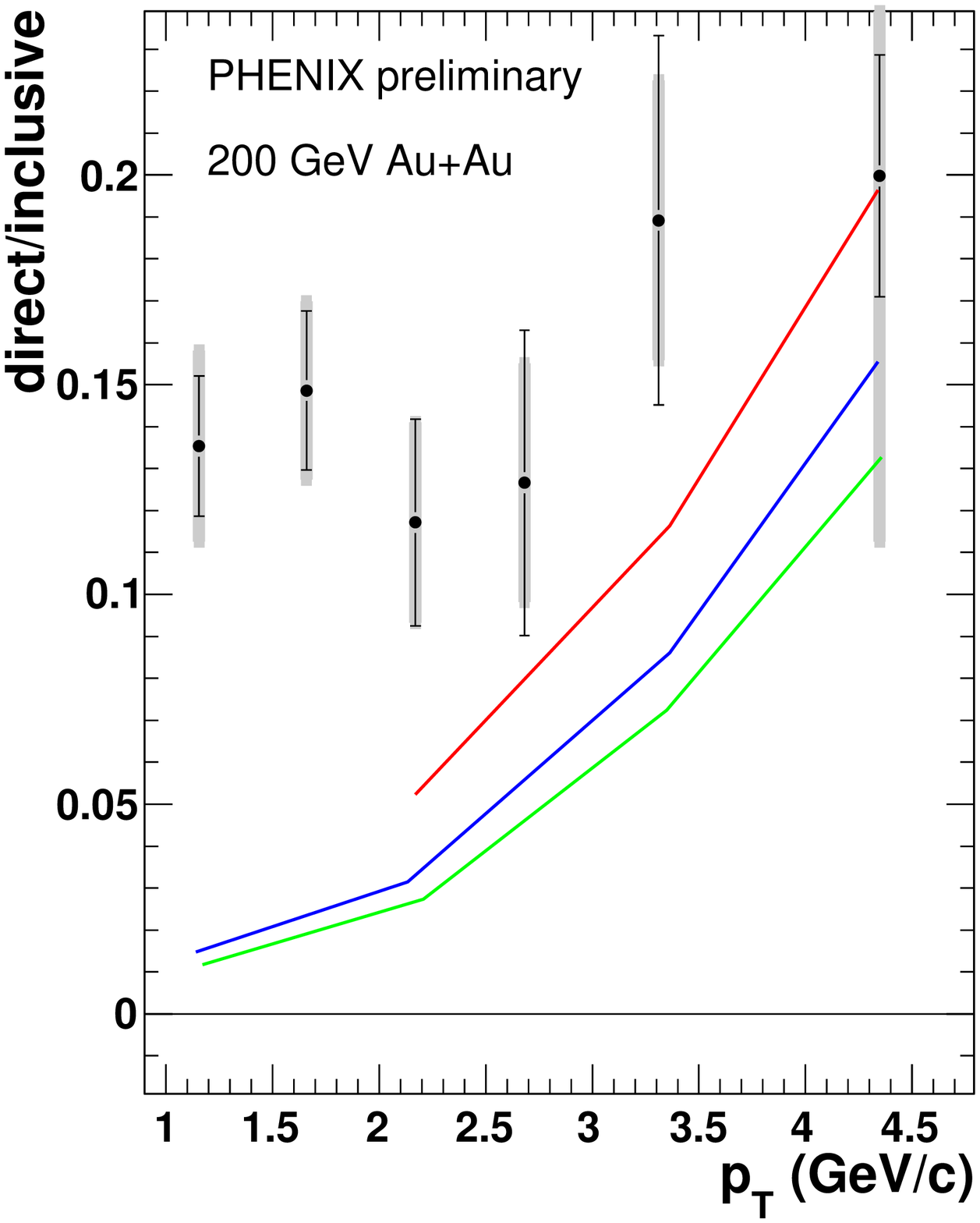}
      \includegraphics[width=0.45\textwidth]{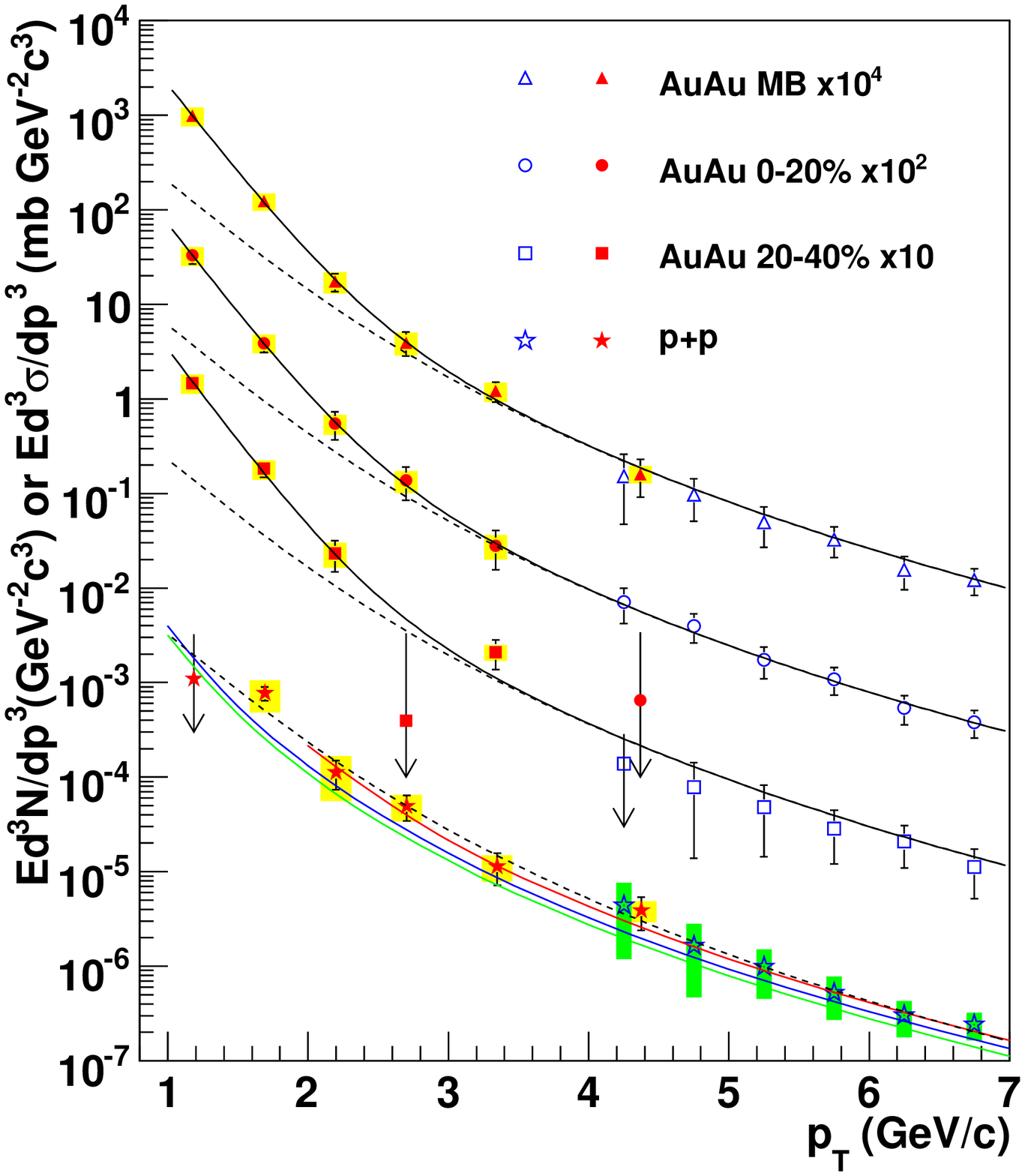}
  \end{center}
  \end{minipage}

  \vspace{-0.5cm}
  \begin{flushleft}
  \begin{minipage}{.48\linewidth}
      \caption{\label{fig:AuAu-rgamma} Ratio of direct-to-virtual photons. 
       The plot is similar to Fig.~\ref{fig:pp-rgamma} but for \AuAu 
       rather than \pp collisions. The pQCD prediction was scalled by 
       the number of binary collisions.} 
  \end{minipage}
  \end{flushleft}
  \vspace{-2.3cm} 
  \begin{flushright}
  \begin{minipage}{.48\linewidth}
      \caption{\label{fig:AuAu-directgamma} 
	Invariant cross section of direct photon production deduced 
	from Fig.~\ref{fig:AuAu-rgamma}. Results from \pp collisions are 
        compared to pQCD calculations, while results for \AuAu are compared 
        to a fit to \pp data scaled by the number of binary collisions.  
        }
  \end{minipage}
  \end{flushright}
  \vspace{-0.3cm}
\end{figure} 

I now turn the attention to the contributions from hard scattering processes. 
Due to the limited statistical accuracy not much can be said about the
contribution from charm production. The data shown in 
Fig.~\ref{fig:AuAu-mass} are consistent with the extrapolation of 
the \pp cross-section using binary scaling, which is labeled $c\bar{c}$ 
(PYTHIA) in Fig.~\ref{fig:AuAu-mass}. However, data on charm production 
\cite{ppg066} indicate a large degree of interaction of charm quarks with 
the medium in central \AuAu collisions, which must also modify the angular 
correlation between $c$ and $\bar{c}$. If the momentum direction of 
the $c$ and $\bar{c}$ quarks would randomized in the medium the data should 
be compared to the curve labeled $c\bar{c}$ 
(random correlation), which leaves room for other contributions above 1 \gev2.

Interesting results emerge when looking at the kinematic region 
of $m<<p_T$, where direct virtual photon emission 
was observed in \pp collisions. Details of the analysis, in particular 
a discussion of the validity of extrapolating virtual photons back to 
the real photon point at mass equal zero, can be found in 
\cite{ppg086,Akiba}. The final result for the direct-to-inclusive 
virtual photon ratio, shown 
in Fig.~\ref{fig:AuAu-rgamma}, indicates a significant excess beyond the 
expectation from initial state $qg$ Compton scattering. 
Fig.~\ref{fig:AuAu-directgamma} shows the direct photon spectrum resulting 
from Fig.~\ref{fig:AuAu-rgamma}. Here the virtual photon data were 
extrapolate to the real photon point, which in this case means 
setting the direct-to-inclusive ratio for virtual photons equal to the one 
for real photons. The data are consistent with results obtained from 
calorimeter measurements at higher \pt in the same experiment. While 
\pp data follow the expectation from pQCD an excess is found in Au$+$Au. 
The inverse slope of the excess, after subtracting the scaled \pp data 
are $T_{eff}\sim220$ MeV. This effective temperature can be interpreted as 
limit on the initial temperature of the reaction volume. Work to confirm 
these results with data from real photons is underway and has produced 
consistent upper limits.  

\section{Model comparison and discussion}\label{sec:model}

Much work has gone into modeling dilepton and photon production from the 
fireball created in heavy ion collisions at SPS energies \cite{spsreview}. 
After some 20 years of experimental and theoretical development the following 
picture has emerged. Electromagnetic radiation is predominantly emitted from
hadron-hadron collisions at times when the spectral function of mesons are 
significantly broadened compared to vacuum. The colliding hadrons participate 
in the collective expansion and consequently the dileptons show radial flow.  
While emission at the partonic level is also observed it does not contribute
significantly to the over all yield \cite{spsreview}.

\begin{figure}[h!]
  \vspace{-0.4cm}
\begin{flushleft}
  \begin{minipage}{1.\linewidth}
  \begin{center}
      \includegraphics[width=0.55\textwidth]{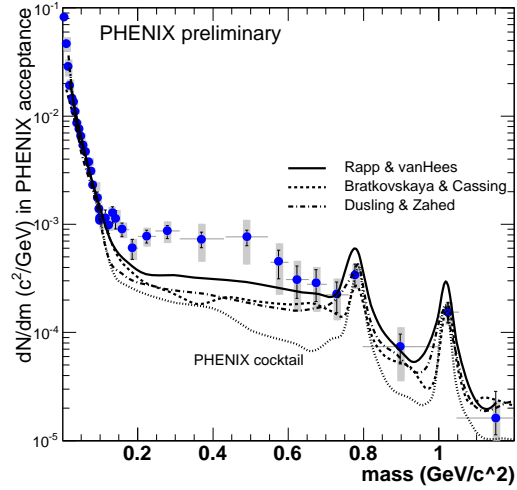}
  \end{center}
  \end{minipage}
\end{flushleft}

  \vspace{-.4cm}
  \begin{minipage}{1.\linewidth}
      \caption{\label{fig:theory-mass} Comparison of \AuAu data to
	various models that successfully describe data from lower 
	energies. Only calculation assuming broadening of the meson masses 
        in the medium are shown.}
  \end{minipage}
  \vspace{-.1cm}
\end{figure}

The same sources must also contribute to dilepton production 
at RHIC energies. In Fig.~\ref{fig:theory-mass} models, which describe
SPS data, are confronted with dilepton mass distribution 
\cite{rapp,cassing,zahed}. 
Evidently, the predicted yield underestimates the data 
and these models are insufficient to describe RHIC data. Most likely 
the contributions from the partonic phase must be more significant than
realized. At closer inspection it turns out that the model 
calculations shown in Fig.~\ref{fig:theory-mass} include only 
$q\bar{q}$-annihilation but not the contribution from $qg$ Compton 
scattering in the medium. 

\begin{figure}[b!]
  \vspace{-0.4cm}
\begin{flushleft}
  \begin{minipage}{1.\linewidth}
  \begin{center}
      \includegraphics[width=0.57\textwidth]{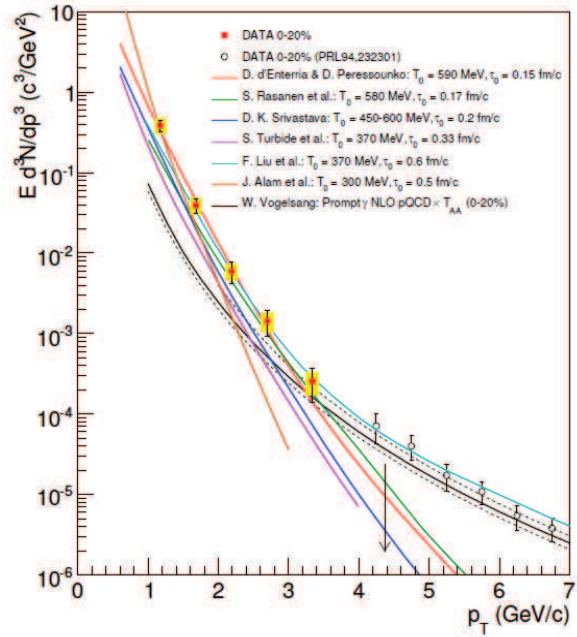}
  \end{center}
  \end{minipage}
\end{flushleft}

  \vspace{-1.2cm}
  \begin{flushright}
  \begin{minipage}{1.\linewidth}
      \caption{\label{fig:theory-gamma} 
        Comparison of \AuAu data on direct photon emission with theoretical 
        models. }
  \end{minipage}
  \end{flushright}

  \vspace{-.1cm}
\end{figure}

Model calculations of thermal photon emission based on perturbative QCD
show that $qg$ Compton scattering is very important \cite{Akiba}. 
In a QGP the gluon densities might be so large that $qg$ Compton 
scattering outshines $q\bar{q}$-annihilation and the perturbative approach 
may be insufficient all together. In Fig.~\ref{fig:theory-gamma} various 
models are compared with real photon data \cite{david}. 
All models agree roughly with the observed slope. The initial 
temperature in the models varies from 300 to 600 MeV depending on the 
initial conditions, in particular the formation time. However, there are large 
differences in the predicted cross sections, which need to be addressed in 
future work. 

\begin{figure}
  \begin{minipage}{1.\linewidth}
  \begin{center}
      \includegraphics[width=1.\textwidth]{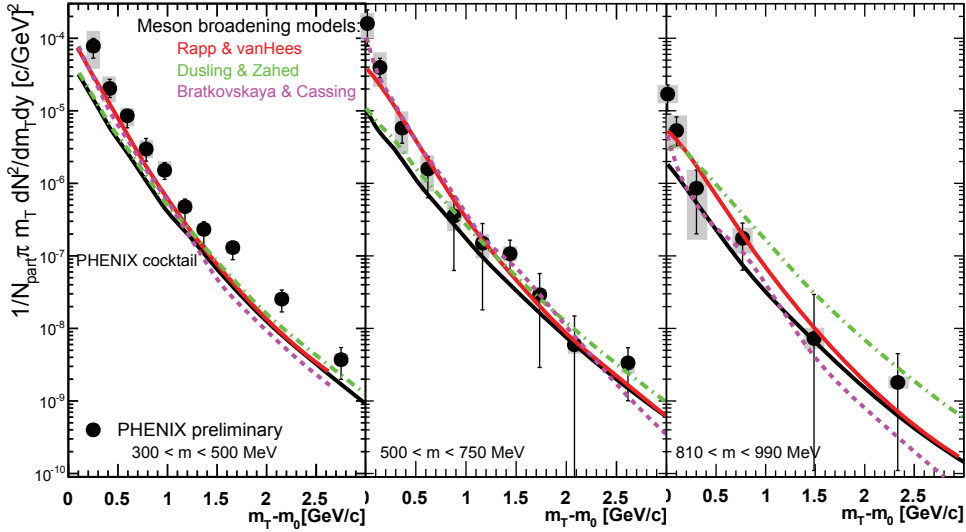}
  \end{center}
  \end{minipage}
  \vspace{-0.5cm}
  \begin{flushleft}
  \begin{minipage}{1.\linewidth}
      \caption{\label{fig:theory-mt} Comparison of \AuAu data to
	various models that successfully describe data from lower 
	energies. The labeling scheme is identical to Fig.~10.}
  \end{minipage}
  \end{flushleft} 
  \vspace{0.1cm}

\end{figure}

Last I return to dilepton spectra, but in a double differential representation 
in mass and \mT. In Fig.~\ref{fig:theory-mt} the same calculations already 
shown in the mass projection (Fig.~\ref{fig:theory-mass}) are compared to 
the \mt spectra. In the mass range from 500 to 750 \mev2 the models describe 
the data except for the lowest \mT, as expected from 
Fig.~\ref{fig:theory-mass}. At lower masses the models fall short at all \mT. 
It is interesting to point out that the data 
above \mt$\sim$1 \gev2 were used to determine the direct photon yield. If 
$qg$ Compton scattering would be consistently included in the models it would 
fill some of the difference at lower \mT, though probably at the lowest
\mt it would not account for all data. Another important consequence may be
that the models then would likely overestimate the yield in mass 
range 500 to 700 \mev2. Obviously more complete theoretical calculations 
are required. In the highest shown mass range the models predict widely 
different yields, most likely due to different contributions from hard 
scattering processes. Again this needs to be looked into.
  
To conclude this section I want to underline the importance of consistent 
model calculations that not only have a realistic description of the space 
time evolution and hadron spectra, but that also describe dilepton and 
photon emission simultaneously.

\section{Summary and outlook}\label{sec:summary}

Dilepton data analysis of the PHENIX experiment has matured significantly over 
the past years. Experimental issues, most importantly the subtraction of 
pair background, are under control. Data from \pp collisions establish a 
precise reference to detect medium effects in \AuAu collisions. PHENIX
has discovered such effects, namely a significant enhancement of dilepton 
production in mostly central \AuAu collisions. The enhancement covers 
the mass range from roughly 100 MeV to 750 MeV and is focused at low \mT. 
Most strikingly it exhibits a prominent soft contribution with 
\teff$\sim$120 MeV, which is independent of mass over the accessible range. 
PHENIX has also presented a first measurement of thermal photons 
indicating an initial temperature larger than 220 MeV. 

Though much work on theoretical models is needed to make final conclusions 
some firm conclusions can already be made. Thermal radiation from the 
hadronic phase, mostly from $\pi\pi$-annihilation with collision broadening
has emerged as the main contribution to dilepton enhancement observed at the 
SPS \cite{spsreview}. These contributions that obviously must exist also at 
RHIC energies remain insufficient to explain the PHENIX data. More 
complete calculations of dilepton radiation from the partonic phase need to 
be made. Model calculations of thermal real photon production is consistent
with the data and hint towards initial temperatures larger than 300 MeV. 
However, big differences between various calculations remain. 

PHENIX has just started their dilepton program, more data 
from \CuCu, \pp, and \dAu collisions are on tape and are being analyzed.   
During RHIC Run 9 the collaboration successfully commissioned the hadron blind 
detector (HBD) and we can look forward to precision \AuAu data from 
PHENIX from Run 10.  


\section*{Acknowledgments} 
Particular thanks go to Ralf Rapp and Yasuyuki Akiba for many discussions 
during the conference.

\end{document}